\documentclass{article}
\usepackage{graphicx}
\usepackage{epstopdf, epsfig}
\usepackage{amsmath}
\usepackage{amssymb}
\usepackage{graphicx}
\usepackage{hyperref}
\usepackage{cleveref}
\usepackage{color}
\usepackage{float}
\usepackage{array}
\usepackage{booktabs}
\usepackage{subcaption}
\usepackage{cite}
\usepackage{float}
\usepackage[export]{adjustbox}
\usepackage[a4paper,left=35mm,right=35mm]{geometry}
\usepackage{amsfonts}
\usepackage{booktabs}
\usepackage{siunitx}

\usepackage[round]{natbib}
\definecolor{blue1}{RGB}{24, 47, 121}
\definecolor{blue2}{RGB}{31, 60, 157}
\definecolor{blue3}{RGB}{39, 75, 193}
\definecolor{blue4}{RGB}{60, 96, 215}
\definecolor{blue5}{RGB}{95, 125, 222}

\definecolor{darkred}{RGB}{166, 42, 23}
\definecolor{blue}{RGB}{48, 117, 181}
\definecolor{green}{RGB}{84, 174, 50}
\definecolor{orange}{RGB}{227, 121, 46}

\newcommand{\vect}[1]{\boldsymbol {#1}}     
\renewcommand{\u}{\vect{u}}                 
\newcommand{\pp}{\partial}                  
\newcommand{\dd}[2]{\frac{\pp #1}{\pp #2}}  
\newcommand{\mathsfbi}[1]{\mathbf{#1}}
\newcommand{\Rey}{\textit{Re}}                 

\title{CubismAMR - A C$++$ library for Distributed Block-Structured Adaptive Mesh Refinement}
\author{Michail Chatzimanolakis, Pascal Weber, \\Fabian Wermelinger, Petros Koumoutsakos}

\begin{document}

\maketitle

\begin{abstract}
We present CubismAMR, a C$++$ library for distributed simulations with block-structured grids and Adaptive Mesh Refinement. A numerical method to solve the incompressible Navier-Stokes equations is proposed, that comes with a novel approach of solving the pressure Poisson equation on an adaptively refined grid. Validation and verification results for the method are presented, for the flow past an impulsively started cylinder.
\end{abstract}

\section{Introduction} \label{sec:Introduction}
Fluid flow simulations involve a multitude of scales and quickly exhaust the capabilities of simulations that employ uniform grids. Adaptive Mesh Refinement (AMR) tackles the issue by dynamically adapting the grid to capture the emerging structures in a flow field. Simulations using AMR that efficiently exploit modern, massively parallel computer architectures are a subject of ongoing investigations. Existing AMR software that targets large-scale simulations include AMReX~\citep{AMReX}, Chombo~\citep{Chombo}, SAMRAI~\citep{SAMRAI2001}, and Basilisk~\citep{vanhooft2018}. Patch-based AMR dynamically groups cells to form rectangular patches in AMReX, Chombo, and SAMRAI. In contrast, Basilisk uses cell-based AMR by organizing the cells in an octree. The AMR framework presented here is based on the Gordon-Bell-Prize winning Cubism library that partitions the simulation domain into cubes and distributes them to multiple compute nodes. These cubes are divided into blocks for cache-optimized thread-level parallelism~\citep{Hejazialhosseini2012,Rossinelli2013,Hadjidoukas2015}. CubismAMR organizes this block-structure into an octree to perform compression and refinement in different regions of space. 

In the present paper we present a numerical method for solving the incompressible Navier-Stokes equations through the use of CubismAMR. The paper is structured as follows: \cref{sec:Numerical Method}  describes the governing equations and their discretization. \Cref{sec:AMR} presents the AMR algorithm we employ, while \cref{sec:Verification and Validation} shows the validation and verification of our method.

\section{Governing Equations}\label{sec:Numerical Method}
We solve the incompressible Navier-Stokes equations in the velocity-pressure formulation
\begin{equation}\label{eq:Incompressible-Navier-Stokes}
\begin{split}
&\mathbf{\nabla}\cdot\u=0\,,\\
&\dd{\u}{t}+(\u \cdot \mathbf{\nabla})\u=-\frac{1}{\rho}\nabla p+\nu\mathbf{\nabla}^2 \u,
\end{split}
\end{equation}
where $\u,p$ and $\nu$ are the fluid velocity, pressure and kinematic viscosity.  The no-slip boundary condition is enforced on the surface of moving bodies with a prescribed velocity $\u^{s}$ through the penalisation approach \citep{Angot1999,Coquerelle2008,Gazzola2011}. The penalisation method augments the governing Navier-Stokes equations by a penalty term $\lambda \chi (\u^{s}-\u)$ where $\chi$ is a characteristic function with values $\chi=1$ inside the cylinder and $\chi=0$ outside. As the penalisation coefficient, denoted by $\lambda$, tends to infinity, \cref{eq:Incompressible-Navier-Stokes} converge to the incompressible Navier-Stokes equations. \cite{Ueda2021} demonstrate that the tangential and slip velocities on the surface of the solid body are in the order of $\lambda^{-1/2}$ and $\lambda^{-1}$ respectively.%
\subsection{Discretisation}
\Cref{eq:Incompressible-Navier-Stokes} is solved using a pressure projection method \citep{Chorin1968} and a second-order time-stepping scheme. These methods are adapted to take into account the penalisation force terms through a split-step algorithm. Quantities measured at time $t=n \Delta t$ will have a superscript $n$. Given the numerical solution at timestep $n$, an intermediate velocity $\u^*$ is computed with the midpoint method
\begin{equation}
\begin{split}
    & \u^{n+1/2}=\u^n+\frac{1}{2}\Delta t\mathbf{F}(\u^n)\,,\\
	&\u^{*}= \u^n + \Delta t \mathbf{F}(\u^{n+1/2})\,,\\
\end{split}
\end{equation}
where $\mathbf{F}(\u)=-(\u \cdot \mathbf{\nabla})\u+\nu\mathbf{\nabla}^2 \u$. Diffusion terms present in $\mathbf{F}(\u)$ are discretised with centered second-order finite differences and advection terms are handled with an upwind fifth-order WENO scheme \citep{Shu1999}. 

A second intermediate velocity is computed by performing an implicit penalisation step
\begin{equation}
	\u^{**}=\u^{*}+\lambda\Delta t \chi^{n+1}(\u^{s,n+1}-\u^{**})\,,
	\label{eq:penal}
\end{equation}
where the discontinuous characteristic function $\chi^{n+1}$ is discretised by using a second-order accurate mollification, proposed by \cite{Towers2009}. \Cref{eq:penal} is stable for any $\lambda>0$, allowing for arbitrarily large values for the penalisation coefficient. The timestep is concluded with a pressure correction step
\begin{equation}
	\u^{n+1}=\u^{**}-\Delta t\nabla p^{n+1}\,,
\end{equation}
where the pressure field is obtained by solving the Poisson equation 
\begin{equation} \label{eq:poisson01}
	\nabla^2 p^{n+1}=\frac{1}{\Delta t}\nabla\cdot\mathbf{u}^{**}\,.
\end{equation}

The equations are solved using block-structured grids. 
The present work relies on the \textsc{CubismAMR} software that is an adaptive version of the \textsc{Cubism} library, which partitions the simulation domain into cubes of uniform resolution that are distributed to multiple compute nodes. These cubes are further divided into blocks for cache-optimised parallelism~\citep{Hejazialhosseini2012,Rossinelli2013,Hadjidoukas2015}. \textsc{CubismAMR} organizes these blocks in an octree data structure (for three-dimensional simulations) or a quadtree data structure (for two-dimensional simulations), which allows for grid refinement or compression in different regions. In contrast to uniform grids or body fitted meshes, this allows to dynamically adopting the grid to capture the emerging structures in a flow field. \textsc{CubismAMR} fully supports two- and three-dimensional distributed simulations. What follows in \cref{sec:AMR} only describes the two-dimensional case, which is relevant to the simulations of this paper.

\subsection{Force computation}
The total force acting on an solid body is
\begin{equation}\label{eq:force_raw}
    \vect{F}= \int\limits_{\partial\Omega_{s}} (2 \mu \mathsfbi{D} \cdot \boldsymbol{n} - p\boldsymbol{n}) \mathrm{d} S\,,
\end{equation}
where $\mathrm{d} S$ denotes the infinitesimal surface element with normal $\boldsymbol{n}$, $\mu$ is the dynamic viscosity, $\Omega_s \in \Omega$ is the subset of the simulation domain occupied by the body and $\mathsfbi{D}=\frac{1}{2}\left(\nabla \u+(\nabla \u)^\top\right)$ the strain-rate tensor. The first term corresponds to viscous forces, and the second to pressure-induced forces.

Following \cite{TOWERS2008} the surface integral in \cref{eq:force_raw} is expressed as a volume integral 
\begin{equation}\label{eq:force_volume}
    \boldsymbol{F}= \int\limits_{\Omega} (2 \mu \mathsfbi{D} \cdot \boldsymbol{n} - p\boldsymbol{n}) \delta (\operatorname{S_d})\;\mathrm{d} \Omega\,,
\end{equation}
where $\delta$ is the Dirac delta and $\operatorname{S_d}$ is the signed distance function from the surface of the body to any point in $\Omega$.
Since $\chi^{(s)}=H(\operatorname{S_d})$, where $H$ is the Heaviside function, we find that $\delta(\operatorname{S_d}) = \dd{\chi}{\mathbf{n}} = \nabla \chi \cdot \boldsymbol n$ where the normal vector is computed from the signed distance function as $\boldsymbol{n}=\frac{\nabla \operatorname{S_d}}{|\nabla \operatorname{S_d}|}\big|_{\partial\Omega_s}$. Thus, the total force is computed from
\begin{equation}\label{eq:force_used}
    \boldsymbol{F}= \int\limits_{\Omega} (2 \mu \mathsfbi{D} \cdot \boldsymbol{n} - p\boldsymbol{n}) (\nabla \chi^{(s)} \cdot \boldsymbol{n})\;\mathrm{d} \Omega\,.
\end{equation}
When computing viscous forces, \cite{Verma2017} observed that velocity gradients near walls can be underestimated, when a penalisation method is used. Therefore, they suggested computing the gradients on a “lifted" surface. Here we employ a similar approach: the necessary gradients are computed two grid points away from the surface and are then extrapolated back to it through a second-order Taylor expansion
\begin{equation}\label{eq:force_approx}
\frac{\partial u}{\partial x_j} = \frac{\partial u_{L}}{\partial x_j} + \Delta x_i \frac{\partial^2 u_{L}}{\partial x_j\partial x_i}~,~ j=1,2\,,
\end{equation}
where the subscript $L$ is used to denote a quantity on the lifted surface, $\Delta x_j$ is distance in the $j$ direction of the grid point of the actual surface from the grid point on the lifted surface. All derivatives are computed with second-order one-sided differences, facing away from the wall. From this the drag $F_D$ and drag coefficient $C_D$ can be computed as the projection of the total force onto the direction of motion
\begin{equation}\label{eq:drag}
    F_D=\boldsymbol{F}\cdot\frac{\mathbf{u}^{(s)}}{|\mathbf{u}^{(s)}|}\,, \qquad C_D=\frac{2F_D}{\rho |\mathbf{u}^{(s)}|^2 A}\,,
\end{equation}
where $A$ is a characteristic area and $\rho=1$ is the fluid density.

\section{Block-Structured AMR}\label{sec:AMR}
The grid is composed of square blocks, each with the same number of cells. Each block has a locally uniform resolution that is defined by its level of refinement $l=0,\dots,L-1$ as $h_l = 2^{-l}h_0$, where $h_0$ is the coarsest grid spacing possible. At mesh refinement from level $l$ to level $l+1$ a block is divided into four blocks , whereas mesh compression from level $l$ to $l-1$ is achieved by combining four blocks to one. Blocks that are adjacent are not allowed to differ by more then one refinement level. Therefore, the blocks of the grid are logically arranged in a quadtree data structure. 

Whether a block should be refined or compressed is determined every few timesteps. Our simulations use the vicinity to solid body surfaces and the magnitude of vorticity as criteria for mesh refinement or compression. Additional examples of such criteria could be the magnitude of pressure gradients, or the magnitude of Wavelet detail coefficients \citep{rossinelli2011b,VASILYEV2005}. Following the multiresolution framework by \cite{HARTEN1994,HARTEN1996}, restriction and prolongation operators are used to map values between blocks of different resolutions. Let
\begin{equation}
    f_{i,j}^l=f\left(\frac{h_l}{2}+i h_l,\frac{h_l}{2}+j h_l\right)
\end{equation}
with indices $i,j\in \mathbb{Z}$ denote the value of $f:\mathbb{R}^2 \rightarrow \mathbb R$ at a cell center. The restriction operator $\mathcal{R}$ is used during mesh compression to replace four blocks at level $l+1$ by one block at level $l$. Grid point values at level $l$ are computed by averaging (see \cref{fig:amr2d})
\begin{equation}\label{eq:restriction}
    f^{l}_{i,j,k} =\frac{1}{4} \sum_{I=0,1}\sum_{J=0,1} f^{l+1}_{2i+I,2j+J}+{\cal O}(h_l^2)\,.
\end{equation}
The prolongation operator $\mathcal{I}$ of one block at level $l$ to four blocks at level $l+1$ is defined via a third-order Taylor expansion
\begin{equation}\label{eq:prolongation}
    f^{l+1}_{2i+I,2j+J} =f^{l}_{i,j} + \frac{h_l}{4} \boldsymbol{d} \cdot \nabla f^{l}_{i,j}
    +\frac{h_l^2}{32} (\boldsymbol{d} \cdot \nabla)^2 f^{l}_{i,j}
    +{\cal O}(h_l^3)\,,
\end{equation}
where $\boldsymbol{d} = (2\delta_{I0}-1,2\delta_{J0}-1)\text{ with }I,J\in\{0,1\}$ and $\delta$ is the Kronecker delta. Derivatives are approximated with second-order central finite difference schemes.

\subsection{Coarse-Fine Interfaces}
%
\begin{figure}
	\centering
	\includegraphics[trim=0 0 0 0, clip,width=0.85\linewidth]{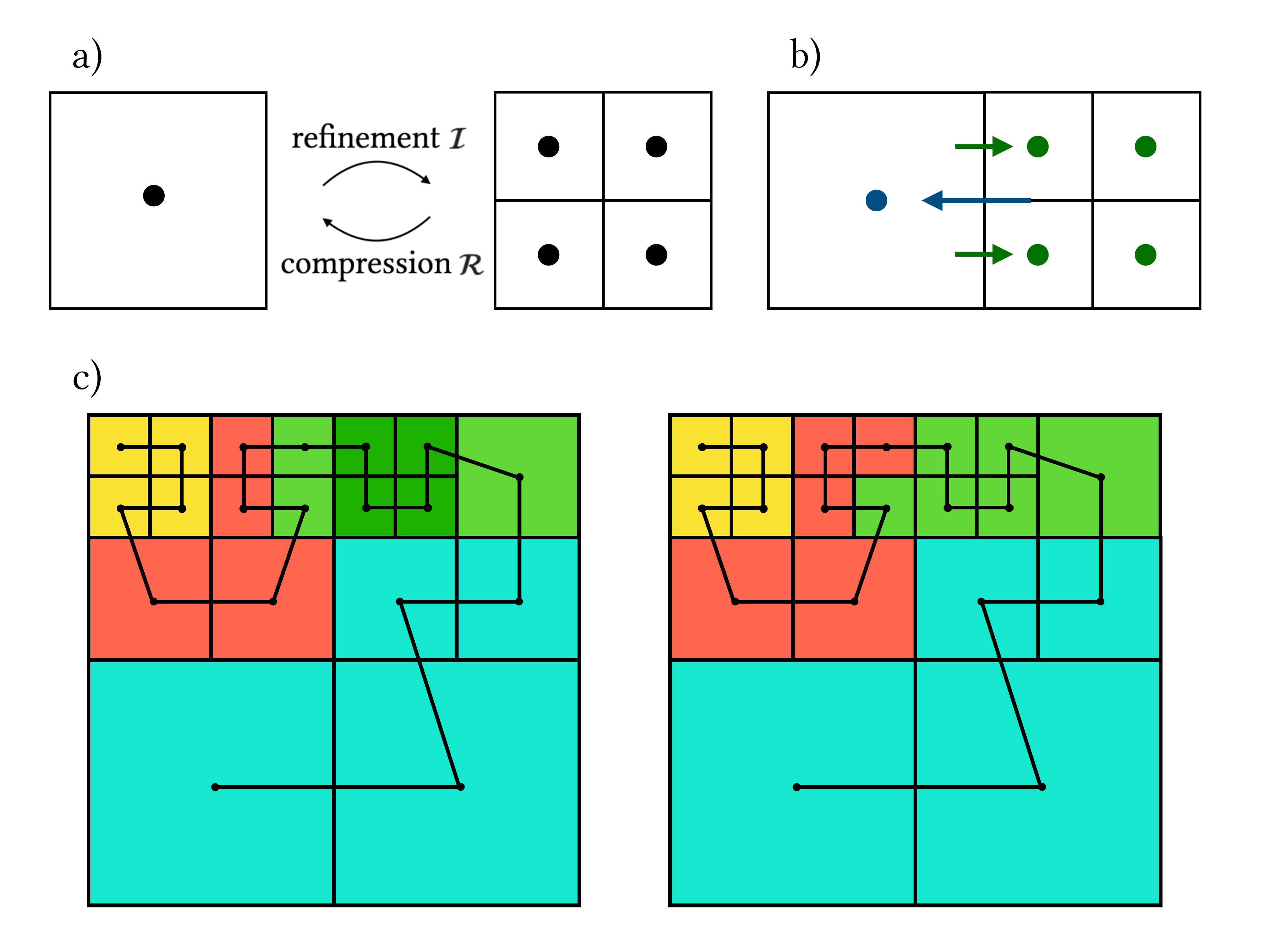}
\caption{Illustration of the AMR. Figure a) illustrates of how cells are refined or compressed.
	b) shows the fluxes at coarse-fine-interface. In order to ensure conservation, the sum of green fluxes is replacing the blue flux. Figures c) illustrate load-balancing using diffusion of work. The black line illustrates the Hilbert curve. The colour represents the process that owns the block. The block in dark green got refined and thus the green process has more work than the other three processes. After diffusion of work the load is again equally distributed.}
	\label{fig:amr2d}
\end{figure}
For the approximation of spatial derivatives, we use finite difference schemes. This is done by creating a frame of uniform resolution around each gridpoint~\citep{rossinelli2015a} through the interpolation of ghost cell values. At the interface between different refinement levels $l$ and $l+1$, this requires the interpolation of ghost cell values from the coarse to the fine level and vice-versa. From a fine to a coarse level this is done by averaging as described in~\cref{eq:restriction}. For the ghost cells that need to be interpolated from a coarse level to a finer one, we use third-order accurate quadratic interpolation as proposed by \cite{Chombo} and \cite{ALMGREN19981}. When used with a second-order finite difference scheme for approximation of first and second derivatives, this guarantees second and first order accuracy respectively.


Whenever needed we follow \cite{Berger1984} and \cite{BERGER1989} and use a second-order accurate conservative discretisation of the divergence operator $\nabla \cdot \boldsymbol{u}$, by expressing it as the sum of fluxes passing through four faces of a cell
\begin{equation}
\begin{split}
    &\nabla \cdot \boldsymbol{u} |_{i,j}^l = \frac{F_{i+1/2,j}^l-F_{i-1/2,j}^l}{h_l}+\frac{G_{i,j+1/2}^l-G_{i,j-1/2}^l}{h_l}+{\cal O}(h_l^2)\,,
\end{split}
\end{equation}
where $F_{i+1/2,j}^l = \frac{u^l_{i+1,j}+u^l_{i,j}}{2}$. Similar expressions are used for the other cell faces. Whenever a cell is next to cells of different resolutions, missing values are interpolated. This results in a non-conservative discretisation of the divergence operator at the interface between two different refinement levels $l$ and $l+1$, as the flux computed for level $l$ need not be equal to the sum of the two fluxes at level $l+1$ that make up the same cell face. The situation is illustrated in \cref{fig:amr2d}. Conservation is achieved by replacing the flux at level $l$ by the sum of the fluxes at level $l+1$. For the fluxes in the +$x$ direction this reads 
\begin{equation}
    F_{i+1/2,j}^l = \frac{1}{2}\left( F_{2i,2j}^{l+1}+F_{2i,2j+1}^{l+1}\right)\,.
\end{equation}
The computation in the other directions is similar.

\subsection{Parallelisation}
\textsc{CubismAMR} is parallelised with the Message Passing Interface (MPI) programming model. The quadtree data structure is traversed by a space-filling Hilbert curve \citep{Hilbert1935}, that assigns each block a unique index. Initially, blocks are distributed to different MPI processes based on their index along the Hilbert curve, the locality property of which guarantees that each process will own blocks that are spatially close to each other. Load-imbalance between different processes, introduced because of mesh compression or refinement, is handled by redistributing work through a one-dimensional diffusion-based scheme. As was first proposed by \cite{CYBENKO1989}, the number of blocks $N_p^{t}$ of process $p$ at timestep $t$ is updated as
\begin{equation}
    N^{t+1}_p = N^t_p + c(N_{p+1}^t - 2N^t_p + N_{p-1}^t)\,.
\end{equation}
$c$ is a user-defined constant, set to $c=0.25$ in the present work. 
This load-balancing scheme limits communication between consecutive processes along the one-dimensional Hilbert curve while gradually redistributing workload. In \cref{fig:amr2d} (bottom) we illustrate the process in two-dimensions. In addition to this, we track the load-imbalance ratio (defined as the ratio between the maximum and the minimum number of blocks any process has). When it exceeds a user-defined threshold (set to 1.1 for all the presented results), all blocks are evenly redistributed to all processes.

\subsection{Solving the Pressure Equation on the Adaptively Refined Mesh}\label{sec:Poisson-Solver}
We write \cref{eq:poisson01} as
\begin{equation} \label{eq:poisson02}
	\nabla^2 \phi=\frac{1}{\Delta t}\nabla\cdot\mathbf{u}^{**}-\nabla^2p^{n}\,,
\end{equation}
where $\phi = p^{n+1}-p^{n}$. \Cref{eq:poisson02} is discretised with a conservative, second-order accurate discretisation of the divergence operator \citep{Martin2008}. All unknown values are concatenated in a vector $\mathbf{\phi} = (\mathbf{\phi}_1 ,\dots, \mathbf{\phi}_B) \in \mathbb{R}^{Bm^2}$, where $\mathbf{\phi}_i\in \mathbb{R}^{m^2}$ corresponds to values found in the $m^2$ grid points owned by block $i$ (assuming a mesh that is composed of $B$ square blocks of locally uniform resolution with $m^2$ grid points each). The arising linear system  $A \boldsymbol{\phi}=\boldsymbol{b}$ is solved by using the preconditioned biconjugate gradient stabilised method \citep{Vorst1992}, with a custom preconditioner. 

The preconditioner is constructed by first defining the Poisson matrix
\begin{equation} 
\mathsfbi M = \begin{bmatrix}
\mathsfbi P & - \mathsfbi I &  \mathsfbi 0 & \mathsfbi 0 & \mathsfbi 0 &\dots & \mathsfbi 0 \\
-\mathsfbi I & \mathsfbi P & - \mathsfbi I & \mathsfbi 0 & \mathsfbi 0 &\dots & \mathsfbi0 \\
 \mathsfbi 0 &-\mathsfbi I & \mathsfbi P & - \mathsfbi I & \mathsfbi 0 &\dots & \mathsfbi 0 \\
 \vdots & \ddots & \ddots & \ddots & \ddots &\ddots & \vdots \\
 \mathsfbi 0 & \dots &  \mathsfbi 0 & -\mathsfbi I & \mathsfbi P & -\mathsfbi I & \mathsfbi 0 \\
 \mathsfbi 0 & \dots & \dots & \mathsfbi 0 & -\mathsfbi I & \mathsfbi P & -\mathsfbi I \\
 \mathsfbi 0 & \mathsfbi 0 & \dots & \dots & \mathsfbi 0& -\mathsfbi I & \mathsfbi P
\end{bmatrix}
,~~
\mathsfbi P = \begin{bmatrix}
4 & -1 &  0 & 0 & 0 &\dots & 0 \\
-1 & 4 & -1 & 0 & 0 &\dots & 0 \\
 0 &-1 & 4 & -1 & 0 &\dots & 0 \\
 \vdots & \ddots & \ddots & \ddots & \ddots &\ddots & \vdots \\
 0 & \dots &  0 & -1 & 4 & -1 & 0 \\
 0 & \dots & \dots & 0 & -1 & 4 & -1 \\
 0 & 0 & \dots & \dots & 0& -1 & 4
\end{bmatrix}\,,
\nonumber
\end{equation}
where $\mathsfbi 0 \in\mathbb{R}^{m\times m}$ is a zero matrix, $\mathsfbi I\in\mathbb{R}^{m\times m}$ is the identity matrix and $\mathsfbi P\in\mathbb{R}^{m\times m}$. This describes the discretisation of the Poisson equation on an $m\times m$ uniform grid with zero Dirichlet boundary conditions, using centered, second-order accurate finite differences. Then, we define the block diagonal matrix 
\begin{equation}
    \mathsfbi A = \text{diag} ( \mathsfbi M,\dots,\mathsfbi M)\,.
    \label{eq:precond01}
\end{equation}
The linear system is preconditioned with $\mathsfbi A^{-1}$. Note that $\mathsfbi M$ is the same for all blocks and it is therefore sufficient to compute its Cholesky decomposition once, and use it whenever it is needed to invert $\mathsfbi A$.
The chosen preconditioner results in the solution of a smaller Poisson equation with homogeneous Dirichlet boundary conditions per block. The (per block) boundary condition $\phi = 0$ is equivalent to assuming $p^{n+1}=p^n$. 

\section{Verification and Validation} \label{sec:Verification and Validation}
We simulate the impulsively started flow around a cylinder, for Reynolds numbers in the range $[550,9~500]$. All the simulations presented in this section use a reference time unit $T=\frac{D}{2U}$, where $D$ is the cylinder diameter and $U$ its velocity. They were performed in a $[0,40D]\times[0,20D]$ domain and the cylinder was placed at $(10D,10D)$. The timestep was controlled by a Courant number of $0.45$ , based on the minimal grid spacing present in the domain. The coarsest grid possible had a resolution of $128\times64$; the finest resolution varied per case. The penalisation coefficient was set to $\lambda=10^7$. Grid refinement and compression were based on vorticity magnitude and the vicinity to the cylinder surface (regions within a distance of $0.1D$ of non-zero values of $\nabla \chi$ were refined to the maximum resolution allowed).
\begin{table}
\begin{minipage}{0.5\linewidth}
\begin{tabular}{cccccc}
& &\multicolumn{2}{c}{Drag coefficient} & \multicolumn{2}{c}{Vorticity} \\
\cmidrule(lr){3-4} \cmidrule(lr){5-6}
 $L$                & $\hat N^L$  &  $\varepsilon_{C_D}^{L}$                                 &  \text{Rate}                  & $\varepsilon_{\omega}^{L}$ & \text{Rate}\\ \midrule 
 4                  &  89         &     0.167                              & 1.01                                 & 0.524  & 1.78\\ 
 5                  & 119         &     0.124                              & 1.97                                 & 0.313  & 2.26\\ 
 6                  & 166         &     0.0469                             & 2.94                                 & 0.125  & 2.63\\ 
 7                  & 242         &     0.0154                             & 2.23                                 & 0.0486 & 2.04\\ 
 8                  & 381         &     0.00779                            & 1.51                                 & 0.0238 & 1.59\\ 
 9                  & 637         &     0                                  & $-$                                  & 0      & $-$  
\end{tabular}
\caption{Verification study errors and convergence rates, impulsively started cylinder at $\Rey=1~000$.} \label{tab:verification}
  \end{minipage}\hfill
  \begin{minipage}{0.45\linewidth}
    \includegraphics[width=\linewidth]{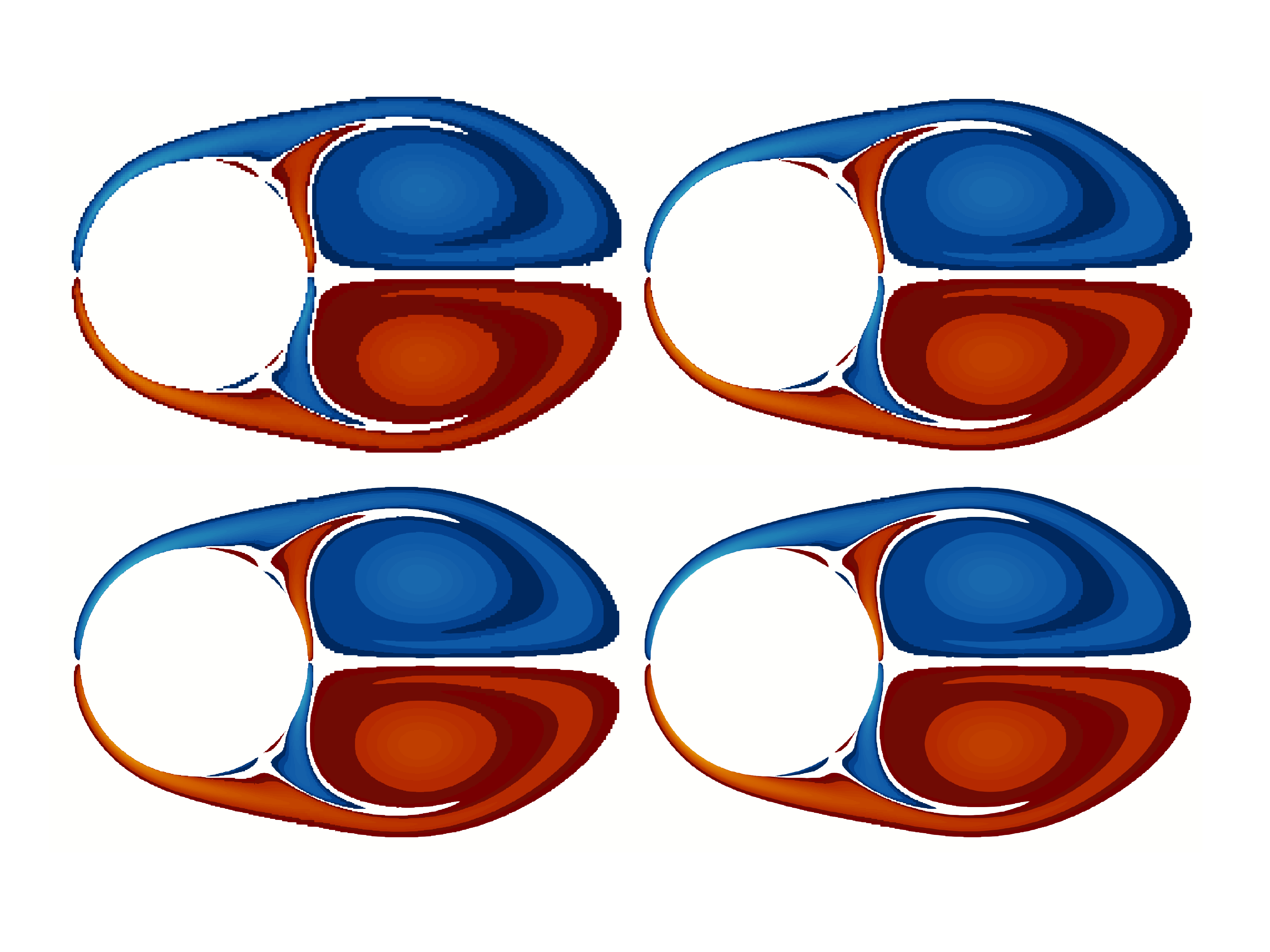}
    \captionof{figure}{Convergence of vorticity field with increasing resolution at $\Rey=1~000$. From top left to bottom right: 6,7,8 and 9 levels of refinement.}\label{fig:space}
  \end{minipage}
  \begin{minipage}{\linewidth}
    \centering
    \includegraphics[width=0.8\linewidth]{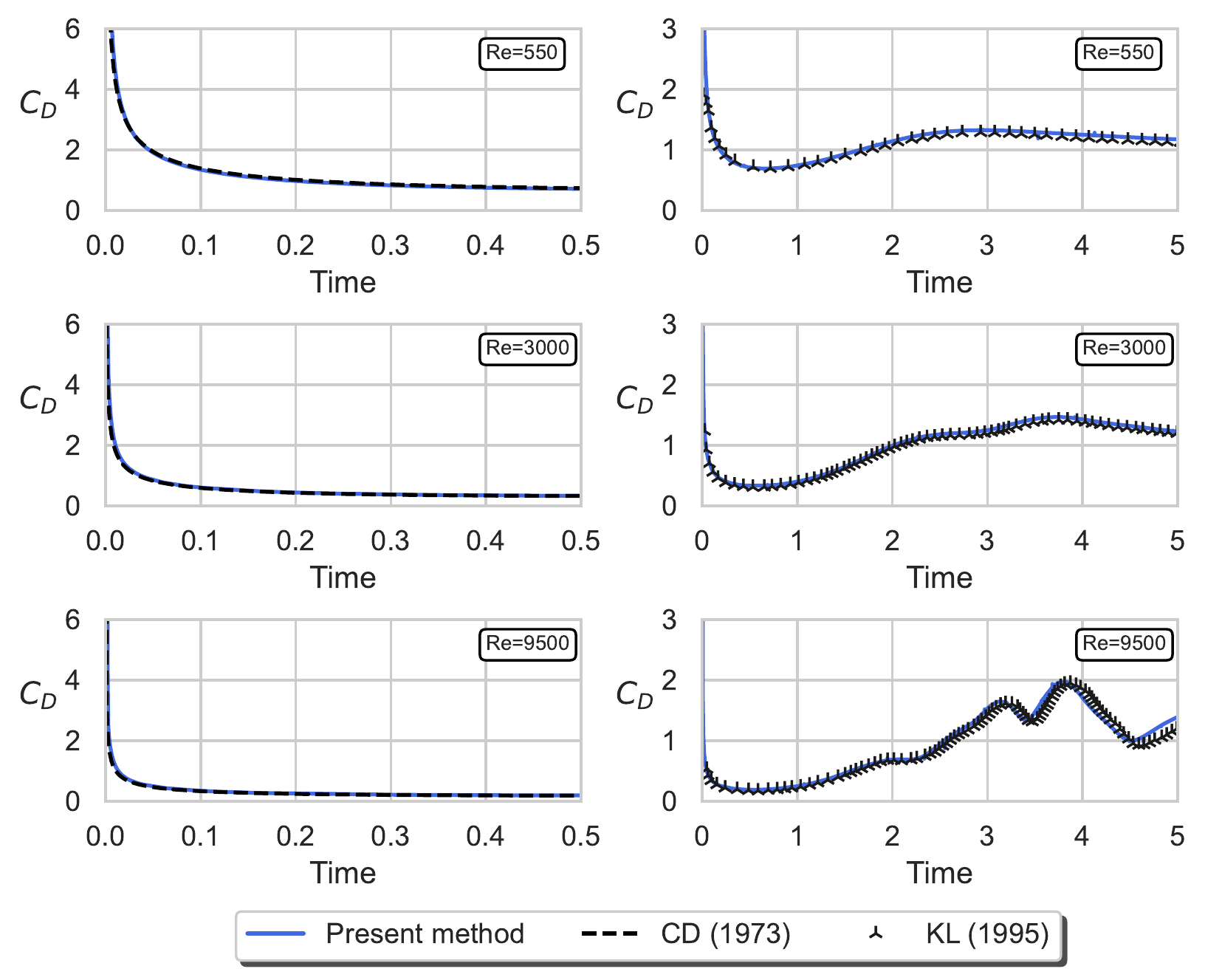}
    \captionof{figure}{Drag coefficient, validation results. Comparison of early drag history with analytical solution by \cite{Collins1973} (left) and with simulations by \cite{Koumoutsakos1995} (right), for various Reynolds numbers.}
    \label{fig:disk_theory}
  \end{minipage}
\end{table}
To verify consistency and convergence, we fix the Reynolds number at $\Rey=1~000$ and perform simulations with increasing resolution, by changing the maximum levels of refinement allowed, denoted by $L$. The simulation with the finest resolution (with $L=L^*=9$ levels) is used as a reference solution for two quantities of interest: the instantaneous cylinder drag coefficient $C_D^L(t)$ and the vorticity field $\omega^L(\tau)$ with $\tau=10T$. For the drag, the error is defined as
\begin{equation}\label{eq:error_drag}
\varepsilon_{C_D}^{L}=\frac{1}{t_{\text{max}}-t_{\text{min}}}\int_{t_{\text{min}}}^{t_{\text{max}}}|C_D^L(t)-C_D^{L^*}(t)|\mathrm{d}t\,,
\end{equation}
with $t_{\text{min}}=0.01T$ and $t_{\text{max}}=10T$ and for the vorticity field as
\begin{equation}\label{eq:error_omega}
\varepsilon_{\omega}^{L}=|\omega^L(\tau)-\omega^{L^*}(\tau)|~~,~~\tau=10T\,.
\end{equation}
Following \cite{Bergdorf2006} we also define the average number of gridpoints per dimension
\begin{equation}\label{eq:cost}
    \hat N^L =\left(\frac{1}{t_{\text{max}}-t_{\text{min}}}\int_{t_{\text{min}}}^{t_{\text{max}}}N^L(t)\right)^{1/2}\,,
\end{equation}
as a measure of the computational cost, where $N^L(t)$ is the instantaneous number of grid points when $L$ refinement levels are used. A summary of the simulations performed for this verification study is shown in \cref{tab:verification}. Convergence of the vorticity field is visualised in \cref{fig:space}, where it can be seen that the solutions for $8$ and $9$ refinement levels are indistinguishable from each other.

To validate our method, we compare our results against the drag coefficient of the impulsively started cylinder at early times as computed analytically by ~\cite{Collins1973} and against simulations by~\cite{Koumoutsakos1995}, for longer times. As shown in \cref{fig:disk_theory}, our results are in excellent agreement with the references they are compared against.
%

\section{Conclusion and Outlook}\label{sec:conclusion}
We presented CubismAMR, a high-performance C$++$ library for distributed simulations with Adaptive Mesh Refinement in two and three dimensions. The library was used with a two-dimensional incompressible flow solver that includes a method of solving the pressure Poisson equation on an adaptively refined mesh. Validation and verification results were shown for the flow past an impulsively started cylinder at several Reynolds numbers that are in excellent agreement with previous simulations and theoretical work. Future work involves making CubismAMR open-source, to provide a reliable tool for making distributed, large-scale AMR simulations.

\appendix

\end{document}